\documentclass[preprint,pra,aps]{revtex4}

\usepackage{graphicx}
\usepackage{dcolumn}   
\usepackage{bm,braket}        
\usepackage{amssymb}   
\usepackage{amsmath}
\usepackage{url,color}
\usepackage{circuitikz}
\usepackage{qcircuit}
\usepackage{layout}
\usepackage{epsfig}
\usepackage{graphicx}
\usepackage{epstopdf}
\usepackage{booktabs}
\usepackage{float,CJK}
\usepackage{titlesec}
\usepackage{multirow}

\def\be{\begin{equation}}
\def\ee{\end{equation}}
\def\ba{\begin{eqnarray}}
\def\ea{\end{eqnarray}}

\begin{document}
\title{Mathematics Is Physical\footnote{This is the English translation of my paper in Chinese [Low. Temp. Phys. Lett. 45(2023)0001]}}
\author{Biao Wu}
\email{wubiao@pku.edu.cn}
\affiliation{International Center for Quantum Materials, School of Physics, Peking University,  Beijing 100871, China}
\affiliation{Wilczek Quantum Center, School of Physics and Astronomy, Shanghai Jiao Tong University, Shanghai 200240, China}
\affiliation{Collaborative Innovation Center of Quantum Matter, Beijing 100871,  China}
\date{\today}

\begin{abstract}
The world of mathematics is often considered abstract, with its symbols, concepts, and topics appearing unrelated to physical objects. However, it is important to recognize that the development of mathematics is fundamentally influenced by a basic fact: mathematicians and computers are physical objects subject to the laws of physics. Through an analysis of the Turing machine, it becomes evident that Turing and his contemporaries overlooked a physical possibility: information carriers can be quantum systems. As a result, computing models like the Turing machine can only process classical information, limiting their computing power. G\"odel's incompleteness theorem highlights the basic fact that mathematicians and computers are made up of finite numbers of atoms and molecules. They can only start with a finite number of axioms, use a finite number of symbols and deduction rules, and arrive at theorems with a finite number of steps. While the number of proofs may be infinite after including all future mathematicians and computers, they must still be enumerable. In contrast, the number of mathematical statements is uncountable, meaning that there will always be mathematical statements that cannot be proved true or false. Just as Landauer claimed that information is physical, mathematics is also physical, limited or empowered by the physical entities that carry it out or embody it.  
\end{abstract}
\maketitle
\section{Introduction}
\label{sec:intro}
Mathematics originated as a practical tool, but it rapidly transcended its practical origins and ventured into the realm of abstract concepts like pure numbers, perfect circles, and lines without width -- objects that do not exist in the physical world. This abstraction empowered mathematics to become more penetrating and precise by liberating the discipline from the constraints of physical objects. Consequently, mathematics has often advanced more swiftly than other branches of science. By the end of the 19th century, the foundational mathematical systems and knowledge on which modern science and technology rely had largely been established, paving the way for unparalleled progress in fields spanning physics to computer science. 

Emboldened by the triumphs of abstraction, mathematicians such as Hilbert, Whitehead, and Russell endeavored to ground pure mathematics on a rigorous, axiomatic foundation. Pure mathematics employs symbols that do not necessarily represent any physical objects, even points without magnitude or lines without width. The objective of this approach was to derive all mathematical theorems from a finite set of axioms using finite rules of logic in a finite number of steps. The ambitious undertaking of Whitehead and Russell, outlined in their seminal work {\it Principia Mathematica}, constitutes a striking exemplar of this effort. Their magnum opus aimed to provide an adamantine basis for the whole of mathematics. 

Notwithstanding the exertions of mathematicians such as Hilbert, Whitehead, and Russell to establish pure mathematics through axiomatization, G\"odel's incompleteness theorem proved in 1931 declared such endeavors futile \cite{godel,nagel}. This theorem asserts that mathematics cannot be reduced to a system of axioms, as there will always be mathematical theorems that are impossible to prove. G\"odel's theorem has sparked considerable mathematical and philosophical discussions. In my view, G\"odel's theorem reflects the fundamental fact that the proof of any mathematical theorem is carried out by finite physical entities, such as mathematicians or machines. These entities can only start with a finite number of assumptions or axioms and derive proofs in finite steps using a finite number of notations and logical rules. Therefore, all mathematical proofs can be expressed as a finite string of symbols and letters. Consequently, the set $\Sigma$ of all mathematical proofs that mathematicians and machines have completed in the past and will complete in the future has a countably infinite number of elements. However, the set $\Omega$ of all mathematical propositions has uncountably infinitely many elements, which means that it is impossible to establish a one-to-one correspondence between the two sets. As a result, there are always mathematical propositions or theorems that are impossible to prove.

In his 1991 article ``Information Is Physical" \cite{Landauer}, Landauer posed a fundamental question: ``Computation is inevitably done with real physical degrees of freedom, obeying the laws of physics, and using parts available in our actual physical universe. How does that restrict the process?" 
G\"odel's incompleteness theorem furnishes an answer to Landauer's query, demonstrating how the underlying physics constrains mathematics and computation. G\"odel's theorem shows that there are always mathematical propositions that cannot be proven by a finite set of axioms, and this implies that there are limits to the computational power of even the most advanced machines. As such, G\"odel's theorem highlights the fundamental nexus between mathematics, computation, and the physical universe in which they are embedded. 

As discussed above, physics can impose constraints on mathematics, but it can also catalyze its progress. A prime example of the latter is the emergence of quantum computers, which demonstrates how insights from physics can empower computation and mathematics \cite{wu,chuang}. 
Mathematicians had thoroughly studied Hilbert spaces and associated linear algebra before the advent of quantum mechanics. However, they did not realize that vectors in Hilbert space could describe a new kind of information, quantum information, which differs fundamentally from classical information in that it cannot be obtained by a single measurement and is not clonable. This understanding was possible only because of the insights furnished by physics. Without quantum mechanics, mathematicians would never have been able to discover quantum information. For quantum computers, which process quantum information, Shor found a much faster algorithm for integer factorization \cite{Shor}. 
This example shows that this new understanding, enabled by physics, can greatly expand the scope and power of mathematics. This illustrates the profound relationship between mathematics and the physical world, and suggests that progress in one field can remarkably benefit the other.

Mathematics is intrinsically connected to physics in two fundamental ways. First, it is performed by physical entities such as human mathematicians or machines, both of which have finite resources. This was famously demonstrated by G\"odel's incompleteness theorem. Second, when mathematical symbols are embodied by physical systems, they can possess properties beyond the purely mathematical. The emergence of quantum information and computing furnishes a compelling example of this. To disregard this nexus is to risk either squandering time in fruitless endeavors or failing to recognize the potential of mathematics and hampering its progress. 
Thus, echoing Landauer's sentiment, we can assert that mathematics is physical. By acknowledging this relationship, we can catalyze further advancements in both mathematics and physics.

In what follows, I will expound on this perspective through an analysis of Turing machines and a discussion of G\"odel's theorem. While I will not provide an in-depth explanation of G\"odel's theorem due to space constraints, I will delve into Cantor's diagonal argument for irrational numbers, which constitutes the most substantial part of the proof of G\"odel's theorem, and Turing's halting problem to demonstrate the connections between mathematics and physics. This paper builds upon and expands the author's previous work \cite{wilczek}.

\begin{figure}
\centering
\includegraphics[width=0.65\linewidth]{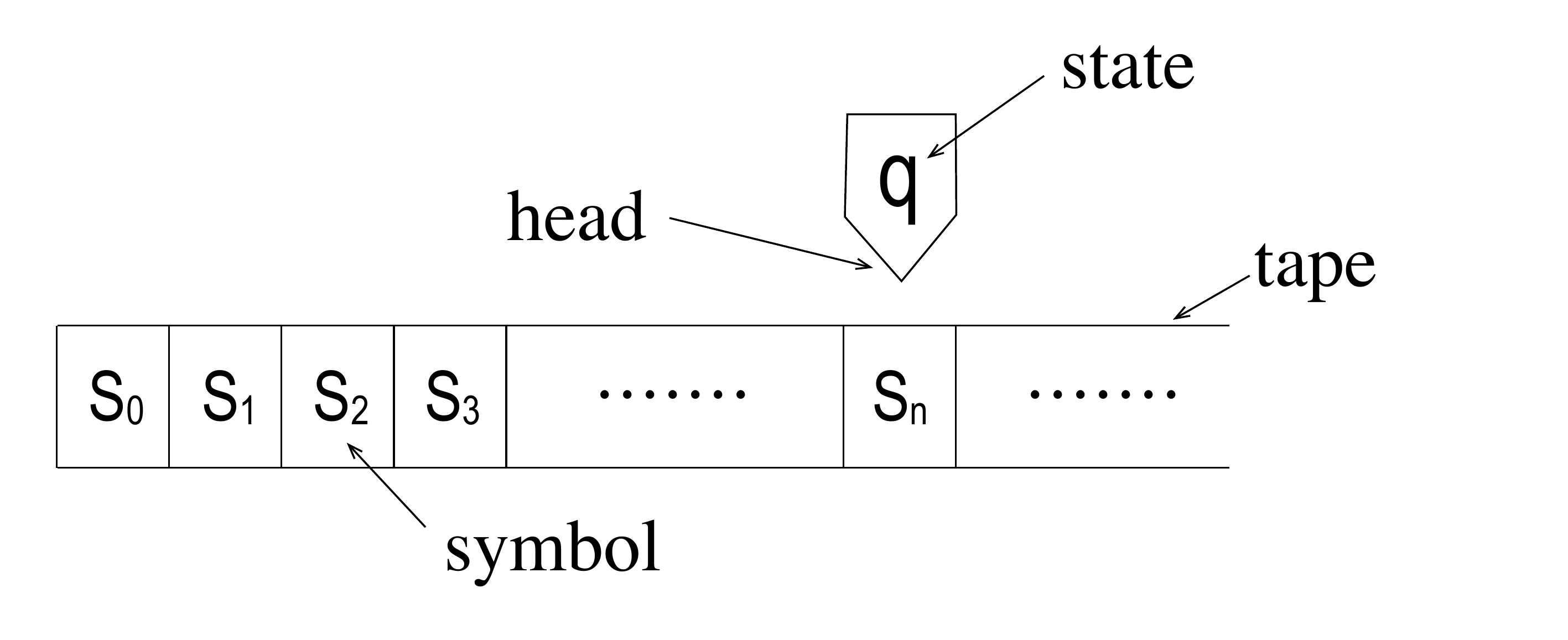}
\caption{Turing machine. }
\label{fig:turing}
\end{figure}

\section{Turing Machine}
\label{sec:turing}
The equation now known as the Schr\"odinger equation \cite{schrod} was published in 1926 and marked the complete establishment of quantum mechanics. Eleven years later, in 1937, Turing introduced the concept of an abstract computer, which is now referred to as a Turing machine \cite{turing}. As depicted in Figure \ref{fig:turing}, a Turing machine consists of a head and a semi-infinitely long strip of tape divided into small squares of equal size. The tape has a finite set of symbols called the alphabet, including a special symbol indicating a blank square. The blank is the only symbol that can appear an infinite number of times on the tape. The machine reads the symbol on the square beneath its head and, based on the symbol and its current state, it rewrites the symbol, changes its state, and moves the head one square left or right, or no movement. A Turing machine is specified by a program $P$ that includes the initial position and state of the head, as well as a finite number of rules in the form of a state table.

\begin{table}[!h]
\caption{State table of a Turing machine}
\label{tab:addtion}
\begin{tabular}{|c|c|c|c|c|c|c|c|c|c|}
\hline
Tape 
& \multicolumn{3}{c|}{State $q_0$}
& \multicolumn{3}{c|}{State $q_1$}
& \multicolumn{3}{c|}{State $q_2$}\\
\cline{2-10}
symbols & write & move & next state & write & move & next state & write & move & next state \\
\hline
0 & 0 & R & Halt & 0 & R & Halt & 0 & R & Halt \\
\hline
1 & 1 & R  & $q_0$& 1 & R &$q_1$ & b & N & Halt\\
\hline
b & 1 & R  & $q_1$ & b & L & $q_2$& b & R & Halt\\
\hline
\end{tabular}
\end{table}

\begin{figure}
\centering
\includegraphics[width=0.7\linewidth]{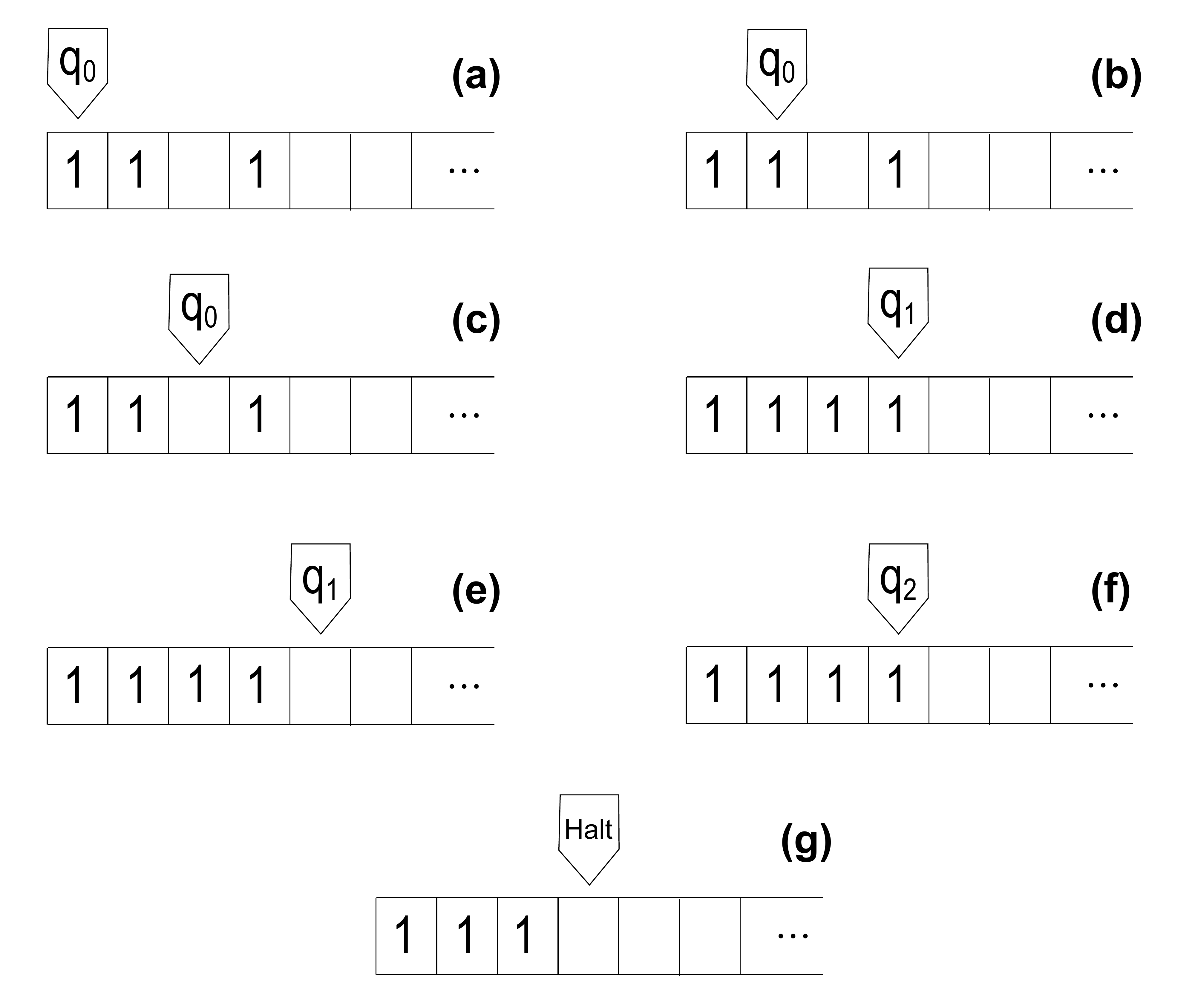}
\caption{An example of a Turing machine. }
\label{fig:turing2}
\end{figure}

Figure \ref{fig:turing2} shows an example of a Turing machine with an initial state of $q_0$ and a state table as shown in Table \ref{tab:addtion}. 
In this table,  the symbol `b' stands for a blank space, `R' signifies a right movement, `L' denotes a left movement, `N' means no movement, and the term ``Halt" indicates that the Turing machine will terminate all operations.  At the start, the tape is blank except for squares 1, 2, and 4, which are set to 1. Following the rules in the table, the Turing machine performs a sequence of operations, gradually evolving from the initial state in Figure \ref{fig:turing2}(a) to the final state in Figure \ref{fig:turing2}(f). This process can be interpreted as the Turing machine computing the sum $2+1=3$ (1 for 1, 11 for 2, 111 for 3) 
or concatenating the strings ``11" and ``1" into the string ``111".

The above introduction illustrates that a Turing machine is an idealized mechanical device employed for computation. Although it was proposed during a time when quantum mechanics was 
well-established, Turing and is contemporaries did not conceive it as a computer related to classical 
physics. In fact, they agreed that Turing machines were universal and could perform any computation -- a concept now known as the Church-Turing thesis. This notion was later strengthened to the widely accepted Church-Turing conjecture, which states that any computation possible on any other computer can also be executed by a Turing machine effectively, meaning that the number of steps required by a Turing machine is proportional to $n^\alpha~(\alpha > 0)$, where $n$ is the number of steps required by the fastest modern computer.

It is now understood that a Turing machine is a classical computer, irrespective of the physical system used to implement it. Envision that there were a world whose mechanics is not Newtonian but Aristotelian, i.e., where the force is proportional to the velocity ($\vec{F}\propto \vec{v}$).  The Turing machines constructed in this world would still be classical computers.  
This is because a Turing machine processes classical information that can be cloned. To elucidate this argument further, we must first understand the fundamental difference between a classical and a quantum dynamical system.

\section{Dynamical Systems}
A dynamical system consists of two parts: the state space and the evolutionary rules. When given an initial condition, the system moves in the state space according to the evolutionary rules, forming a trajectory. The state space and evolution rules differ for classical and quantum particles. For a classical particle, its state space is a six-dimensional phase space, where each point or vector $(\bm{r},\bm{p})$ represents a state of the particle. The evolution of the particle states is governed by the Newtonian equations of motion, 
\begin{equation}
\label{eq:newton}
\frac{d\bm{r}}{dt}=\frac{\bm{p}}{m}\,,~~~~~\frac{d\bm{p}}{dt}=-\nabla V(\bm{r})\,.
\end{equation}
Here $m$ is the mass of the particle and $V(\bm{r})$ is the potential energy. 
On the other hand, for a quantum particle, its state space is Hilbert space, where each vector $\ket{\psi}$ represents a quantum state of the particle. The evolution of the quantum state is governed 
by the Schr\"odinger equation,  which is given by
\begin{equation}
\label{eq:sch}
i\hbar\frac{d}{dt}\ket{\psi}=\hat{H}\ket{\psi}\,,
\end{equation}
where the Hamiltonian operator $\hat{H}=\hat{\bm{p}}^2/(2m)+V(\hat{\bm{r}})$. 
Although the quantum state $\ket{\psi}$ spreads in real space as a wave function, it is a point in Hilbert space, and its time evolution forms a trajectory in Hilbert space. 

A Turing machine is also a dynamical system with a state space consisting of all 
possible states $T$ of the tape. For example, the seven cases of the tape shown in Figure \ref{fig:turing2} correspond to seven points in the state space of a Turing machine. 
The evolutionary rules of a Turing machine are listed in state tables, like the example illustrated in Table \ref{tab:addtion}. 



The equations (\ref{eq:newton}) and (\ref{eq:sch}) reveal that the mathematical forms of classical and quantum dynamics differ markedly. It may be tempting to infer that the crucial difference between these two branches of mechanics stems from their distinct mathematical formulations of dynamical evolution. However, this impression is superficial, not substantive. In fact, the Schr\"odinger equation can be mathematically expressed as a classical Hamiltonian system\cite{Heslot}. 

Let us consider the simplest quantum system, a single spin 1/2, 
for which the Schr\"odinger equation is:
\begin{equation}
i\hbar\frac{d}{dt}
\begin{pmatrix}\phi_1\\ \phi_2
\end{pmatrix}=\begin{pmatrix}H_{11}&H_{12}\\ H_{21}&H_{22}
\end{pmatrix}\begin{pmatrix}\phi_1\\ \phi_2
\end{pmatrix}\,.
\label{eq:two}
\end{equation}
The energy expectation for this system is
\begin{equation}
\label{eq:eng}
{\mathcal H}=\phi_1^*H_{11}\phi_1+\phi_2^*H_{22}\phi_2+\phi_1^*H_{12}\phi_2+\phi_2^*H_{21}\phi_1\,.
\end{equation}
We can treat it as the classical Hamiltonian of two pairs of conjugate variables $\phi_1^*,\phi_1$ and $\phi_2^*,\phi_2$  that satisfy the following Poisson brackets
\begin{equation}
\label{eq:poisson4}
\{\phi_1^*,\phi_1\}=i/\hbar\,,~~~~\{\phi_2^*,\phi_2\}=i/\hbar\,.
\end{equation}
According to the classical Hamiltonian theory, the equations of motion for ${\mathcal H}$ are
\begin{equation}
\frac{d\phi_1}{dt}=\{\phi_1,{\mathcal H}\}\,,~~~\frac{d\phi_2}{dt}=\{\phi_2,{\mathcal H}\}\,,
\end{equation}
which are identical to the equation (\ref{eq:two}). In general,  by choosing a complete set of orthogonal normalized bases, 
we can always write the Schr\"odinger equation (\ref{eq:sch}) in matrix form. Thus for any quantum system, 
its energy expectation can always be expressed in a form similar to that of ${\mathcal H}$ in the equation (\ref{eq:eng}).
Thus by further introducing conjugate variables and Poisson brackets, we can always 
write the Schr\"odinger equation as a classical Hamiltonian equation of motion.
In contrast and complementary to this,  Koopman in 1931 and  von Neumann in 1932 independently discovered that classical dynamics can also be described mathematically in terms of Hilbert spaces and operators. However, we will not delve into this theory here, but instead refer interested readers to their papers \cite{koopman,Neumann}. 

While there may not be an essential mathematical difference between classical and quantum dynamics, it is well 
known that they are physically distinct. So, what accounts for this difference? In the next section, 
I will conduct a detailed comparison between the one-dimensional classical particle and the quantum spin 1/2 systems. 
Through this analysis, I find that the state of a classical dynamical system is classical information, which can be cloned, 
whereas the state of a quantum dynamical system is quantum information, which cannot be cloned. This distinction 
between classical and quantum systems is essential, and it cannot be determined solely through mathematical means.

\section{Unclonable Quantum Information}
In classical mechanics, the state space of a one-dimensional particle is a two-dimensional phase space 
where the state of the particle is represented by a point or vector ${\bm X}=(x,p)$, with $x$ and $p$ being the position and momentum of the particle, respectively. On the other hand, the state space 
of spin 1/2 is a two-dimensional Hilbert space, where the state of the spin is represented by a vector 
\begin{equation}
\ket{\psi}=\phi_1\ket{\uparrow}+\phi_2\ket{\downarrow}\,,
\end{equation}
with $\phi_1$ and $\phi_2$ being the coefficients of the basis vectors $\ket{\uparrow}$ 
and $\ket{\downarrow}$, respectively. 
While both $x,p$ and $\phi_1,\phi_2$ are components of vectors in a given linear space, 
there is a fundamental difference between them. The position $x$ and momentum $p$ are 
directly observable in the ideal case, where instrumentation is infinitely accurate, perturbations 
are infinitely small, and there is no noise. In contrast, $\phi_1$ and $\phi_2$ cannot be measured directly. According to quantum mechanics, a measurement of spin yields two possible outcomes: spin up $\ket{\uparrow}$ with probability $|\phi_1|^2$ and spin down $\ket{\downarrow}$ with probability $|\phi_2|^2$. This illustrates two essential differences between quantum and classical measurement: (1) the outcome of a quantum measurement is uncertain and (2) the measurements are not the components $\phi_1,\phi_2$ of the vectors in Hilbert space, but rather the basis vectors $\ket{\uparrow}$ or $\ket{\downarrow}$. While $\phi_1,\phi_2$ completely determine the state of the spin, their values cannot be obtained from a single measurement. Instead, multiple measurements must be made to obtain $|\phi_1|^2,|\phi_2|^2$ and the phase of the complex $\phi_1,\phi_2$ can only be obtained by taking multiple measurements along either the $x$-axis or the $y$-axis. This is in sharp contrast to classical mechanics where $x$ and $p$ completely determine the state of the particle and their values can be obtained exactly from a single measurement.

The distinctions between quantum and classical systems in relation to measurement are fundamental and cannot be derived solely from mathematics. It is incorrect to argue that $\phi_1$ and $\phi_2$ cannot be measured directly because they are complex numbers. In fact, for one-dimensional classical particles, we can introduce the complex variables $a$ and $a^*$, given by
\begin{equation}
a=\frac{1}{\sqrt{2}}(x-ip)\,,~~~~a^*=\frac{1}{\sqrt{2}}(x+ip)\,.
\end{equation}
They are  a pair of conjugate variables, satisfying the Poisson bracket $\{a,a^*\}=i$. 
Although $a,a^*$  are complex, they are directly measurable since both their real 
and imaginary parts can be measured. 
Similarly, for spin, we can write $\phi_1$ and $\phi_2$ as
\begin{equation}
\phi_1=\phi_{1r}+i\phi_{1i}\,,~~~~\phi_2=\phi_{2r}+i\phi_{2i}\,,
\end{equation}
and substitute them into Eq. (\ref{eq:two}) to obtain a set of equations with only real numbers. 
However, the real variables $\phi_{1r}$, $\phi_{1i}$, $\phi_{2r}$, and $\phi_{2i}$ are still not directly measurable. Thus, the question of whether a quantity is directly measurable depends entirely 
on the physical system being studied and has nothing to do with its mathematical form.

The classical dynamical variables $x,p$ are directly measurable, while the quantum dynamical 
variables $\phi_1,\phi_2$ are not. This distinction is very profound and has important 
physical consequences. In classical measurements, the state of a measuring instrument before the measurement, denoted by $D_0$, is transformed into the state of the instrument after the measurement, denoted by $D_{\bm X}$, such that the complete information about the particle, represented by $\bm X$, is now recorded in the instrument. This process can be expressed as 
\begin{equation}
\label{eq:mc}
{\bm X}\otimes D_0\rightarrow {\bm X}\otimes D_{\bm X}\,.
\end{equation}
It corresponds to a cloning process, where there are now two copies of $\bm X$ in the world. 
In contrast, quantum measurements do not allow for the direct measurement of $\phi_1$ and $\phi_2$, 
and the information about the quantum state $\ket{\psi}$ cannot be recorded exactly 
in a single measurement. Therefore, the process of a quantum measurement  can be expressed as
\begin{equation}
\ket{\psi}\otimes D_0\nrightarrow \ket{\psi}\otimes D_{\ket{\psi}}\otimes D_{\ket{\psi}}\,,
\end{equation}
and quantum measurement cannot be regarded as a cloning process. This is consistent with 
the well known no-cloning theorem, which forbids cloning in quantum systems\cite{Park1970,Wootters1982}. 
This analysis demonstrates that classical particle states can be cloned, while quantum states represented by spin states cannot be cloned.

Different Turing machines implement distinct evolutionary rules, thereby enabling them to perform diverse computations. The Church-Turing thesis asserts that a Turing machine can perform any computable task. Consequently, it is possible to design two distinct Turing machines: one to compute Newton's equation (\ref{eq:newton}) and the other to compute the Schr\"odinger equation (\ref{eq:sch}).  
Therefore, we cannot discern from the evolution rules whether a Turing machine is classical or quantum. This conclusion is in line with our earlier analysis, which emphasized that 
the mathematical form of dynamical equations does not fundamentally distinguish 
between classical and quantum systems.

Let us now consider the state space of a Turing machine, which comprises all possible tape records $T$. For instance, the examples shown in Figure \ref{fig:turing2} correspond to seven points in this space.
We can certainly obtain an accurate picture of the symbols on the tape by looking at it only once, at the same time, the machine's head can read the tape squares in one go. 
This means that we can use equation (\ref{eq:mc}) to describe the process of reading or measuring the tape by replacing ${\bm X}$ with some tape record $t_p$. Therefore, the tape record $t_p$ can be cloned in the same way as the state ${\bm X}=(x,p)$ in classical mechanics. It is for this reason that Turing machines are classical computers and not quantum computers. Note that the Turing machine that computes Schr\"odinger equation (\ref{eq:two}) is still a classical computer, just like a laptop computer that solves Schr\"odinger equation is  a classical computer.

The preceding discussion highlights that classical and quantum information are two distinct types of information. Classical information is clonable, while quantum information is not, as stipulated by the well-known quantum no-cloning theorem\cite{Park1970,Wootters1982}. A computer that directly processes classical information is a classical computer, while a computer that processes quantum information directly is a quantum computer. Turing machines, as devices that process classical information, are classical computers. 
The word ``directly'' is crucial here. As already mentioned, 
both Turing machines and conventional laptops can solve the Schr\"odinger equation (\ref{eq:two}). To accomplish this, they convert the quantum information represented by $\phi_1$ and $\phi_2$ into classical information. 
That is,  the complex variables $\phi_1$ and $\phi_2$ are recorded on the Turing machine's tape 
or stored in the laptop's memory as  four real numbers, which can be read or measured in a single step. As a result, these devices do not process the quantum information directly but instead convert it to classical information. This distinction sets quantum computers apart from classical computers since they process $\phi_1$ and $\phi_2$ directly as a quantum bit state rather than as four real numbers that can be read directly. We will examine this subtle yet profound difference more thoroughly later.

The Turing machine was introduced in 1937, more than a decade after the establishment of quantum mechanics. However, Turing, his contemporaries, and subsequent scientists did not realize that these computational models assumed that the information directly processed by the machines was recorded on media that could be read or copied accurately at once. In everyday life, information is stored on this type of mediums such as paper, film, sound waves, and electromagnetic waves that can be recorded or read out in a single shot. This characteristic aligns perfectly with classical physics since the states of particles and waves in classical systems can, in principle, be observed accurately at once. The association between this property of information and classical physics was so ingrained that it was overlooked for many years. It wasn't until 1980 and 1981 that Soviet mathematician  Manin \cite{Manin} and American physicist  Feynman \cite{Feynman} recognized that Turing machines, by only being able to process classical information, would face fundamental difficulties in dealing with quantum systems.

We can use Turing machines to illustrate the fundamental challenge of simulating 
quantum systems with classical computers. As previously noted, a Turing machine can simulate the motion of a single particle according to the Newton equation (\ref{eq:newton}). However, when we consider a system of multiple particles, the state of the system becomes  
more complex. For example, for a system of two particles, the state is described by four variables $(x_1,p_1;x_2,p_2)$, which represent the positions and momenta of the two particles. When we simulate this system on a Turing machine, the length of the tape needed to record the state of motion doubles compared to the single-particle case. If there are three particles, the state  
is described by six variables $(x_1,p_1;x_2,p_2;x_3,p_3)$, which are the positions and momenta of  the 
three particles. In this case, the length of the tape needed to record the state of motion triples. 
As we add more particles to the system, the complexity grows. For a system of $n$ particles, the state is described by $n$ sets of position and momentum variables $(x_1,p_1;x_2,p_2;\cdots; x_n,p_n)$, 
and the length of data required to record the state of motion on the Turing machine tape grows linearly with $n$.

When computing the Schr\"odinger equation describing a spin (\ref{eq:two}) with a Turing machine, the paper tape needs to record $\phi_1$ and $\phi_2$, which is equivalent to recording four real numbers. If there are two spins, the dimension of the Hilbert space is four and its vector has four components, so recording the state of the system requires eight real numbers. For three spins, the dimension of the Hilbert space is eight, and it takes 16 real numbers to record the state of the system. For a system with $n$ spins, the dimension of the Hilbert space is $2^n$, and it takes $2^{n+1}$ real numbers to record the state of the system. This means that the length of data on the tape of a Turing machine grows exponentially with the number of spins, proportional to $2^n$. Although the tape of a Turing machine is semi-infinite and can accommodate a large amount of data, its computation speed will slow down exponentially for two reasons: first, the time to move each step is fixed, and second, the distance the head has to travel on the tape to read the state of the system increases exponentially with $n$. As a result, the time to read the data alone grows exponentially with $n$. Therefore, Turing machines cannot effectively model multi-body quantum systems due to the exponential divergence of storage space required to simulate many-spin quantum systems on classical computers. This is the fundamental difficulty that Manin and Feynman realized. The cause of the exponential divergence is that the quantum states of $n$-spin systems are processed as classical information rather than as quantum information. If we treat quantum states as quantum information and directly represent the quantum state of the spin system in terms of quantum bits, we only need $n$ quantum bits to ``record" the quantum states of the system of $n$ spins. This approach eliminates the problem of exponential divergence, and quantum computers can be used to efficiently simulate many-body quantum systems.

The above discussion leads to two important conclusions. Firstly, quantum information can be transformed into classical information but at a significant cost. The storage space required by a classical computer, such as a Turing machine, grows exponentially with the size of the system. Secondly, physics can extend the capabilities of mathematics. Quantum computers can efficiently simulate quantum many-body systems and solve some computational problems faster, such as integer factorisation\cite{Shor} and random search\cite{Grover}. It is crucial to emphasize that quantum computers were inspired by physics and  not by mathematics. The mathematics used in quantum computing, such as matrices and Hilbert spaces, existed and were studied intensively long before the establishment of quantum mechanics. However, no one realized that it was possible to build more powerful computer models based on this mathematics. Only through physics could people discover a different type of information, quantum information, which cannot be read by a single measurement and cannot be cloned. 
No matter how one studies Hilbert space mathematically, it is not possible to discover that its vector 
can represent quantum information without new physics. Mathematics is physical in this sense, as it cannot be completely divorced from reality and reduced to a logical relationship between abstract symbols. It is likely that other linear spaces in mathematics will describe new types of information in the future, but without new physics, we cannot know the characteristics of this information or the physical systems that can store it directly. Therefore, the advent of quantum computers shows that physics can give mathematics a richer content and make it more powerful.

The preceding discussion suggests that discovering new kinds of information could lead to the creation of computer models that process this information directly and potentially offer more computational power than quantum computers. In a recent study \cite{HWW}, my colleagues and I proposed a novel type of information that cannot be cloned and part of which  can never be observed. Building on this new information, we developed a new computer model called the Lorentz quantum computer, where some of its logic gates are  Lorentz transformations on the complex domain. Our results demonstrate that this computer model is indeed more powerful than a quantum computer. These findings show that there is still room for innovation in the field of computing and that new types of information can lead to more powerful computer models beyond the limits of quantum computing. By conceiving new kinds of physical information and developing computer models tailored to process them, we can continue to push the boundaries of computation.

\section{Irrational numbers and G\'odel's incompleteness theorem}
\label{sec:godel}
The preceding discussion revealed that physics can expand the scope and capabilities of mathematics, but in contrast, G\"odel's incompleteness theorem, first introduced in 1931, demonstrates that physics can also limit the power of mathematics. The rigorous formulation of G\"odel's theorem states that any formal theory based on a finite number of axioms cannot prove all true propositions about natural numbers, nor can it prove that the system is self-consistent. While the strict formulation of G\"odel's theorem pertains only to propositions about natural numbers, it is generally accepted that there are mathematical propositions in any finite axiomatic system that are unprovable or unfalsifiable. This theorem may seem mysterious, but it reflects the fundamental fact that while mathematical propositions can transcend physics, their proof is accomplished by physical entities, such as humans or machines. Humans and machines can only use a finite number of symbols, methods, and derivation steps, thus limiting the number of mathematical propositions that they can prove or disprove to a countable infinity, equivalent to the number of integers. In contrast, the number of mathematical propositions is uncountably infinite, equal to the number of real numbers. Thus we cannot establish a one-to-one correspondence 
between the two sets, all  mathematical proofs $\Sigma$ and all mathematical propositions $\Omega$. 
As a result,  there are always mathematical propositions that are not provable or falsifiable.
I will not provide a detailed discussion of G\"odel's theorem in this limited space, interested readers can refer to the original paper by G\"odel \cite{godel} or consult the literature \cite{nagel}. Instead I will 
introduce Cantor's diagonal arguments for irrational numbers, which forms 
the basis of the proof of G\"odel's theorem.

In 1891, Georg Cantor, a German mathematician, proved a fundamental result in the theory of sets - that there are more real numbers than there are integers. This conclusion was significant, but equally important was the method Cantor used to prove it. This diagonal method of argument, which he employed in his proof, became highly influential and formed the basis of G\"odel's theorem.

To understand the method, note that the set of real numbers in the open interval $(0,1)$ is equivalent in size to all real numbers. Therefore, we can restrict our attention to the real numbers in $(0,1)$. Cantor used proof by contradiction to show that the set of real numbers in $(0,1)$ is uncountable.

He began by assuming that there is a one-to-one correspondence between the set of real numbers in $(0,1)$ and the set of integers. This implies that we can label all the real numbers in $(0,1)$ with integers, say $r_1, r_2, r_3, \ldots, r_n, \ldots$. We can represent each of these real numbers in binary form by ignoring the zeros and decimal points, and keeping only the digits after the decimal point. This gives us an infinite matrix where the $n$th row corresponds to the binary expansion of $r_n$ as follows
\begin{equation}
\label{eq:cantorm}
\begin{array}{l}
r_1=\\r_2=\\r_3=\\r_4=\\r_5=\\r_6=\\r_7=\\r_8=\\r_9=\\ \vdots
\end{array}
\begin{bmatrix}
\underline{0}&1&1&0&1&0&1&0&0& \cdots\\ 
1&\underline{0}&1&1&1&0&1&1&0& \cdots\\
0&0&\underline{1}&1&1&1&0&0&0& \cdots\\ 
0&1&1&\underline{1}&0&0&1&1&0& \cdots\\ 
1&0&1&0&\underline{0}&1&0&1&0& \cdots\\ 
0&0&1&0&0&\underline{1}&0&0&0& \cdots\\ 
0&1&1&1&0&1&\underline{0}&0&0& \cdots\\ 
1&1&1&1&0&1&0&\underline{1}&0& \cdots\\ 
1&1&1&0&0&1&1&1&\underline{1}& \cdots\\ 
\vdots &\vdots &\vdots &\vdots &\vdots &\vdots &\vdots &\vdots &\vdots &\ddots
\end{bmatrix}\,.
\end{equation}

We can obtain a new real number $\mathfrak{r}$ by taking the diagonal elements of the matrix 
and flipping each of them, replacing 0's with 1's and 1's with 0's. For the matrix in Eq.(\ref{eq:cantorm}), the new number is
\begin{equation}
\mathfrak{r}=0.110010100\cdots\,.
\end{equation}
The first digit of $\mathfrak{r}$ is the opposite of the first digit of $r_1$, the second digit is the opposite of the second digit of $r_2$, and so on. The resulting number $\mathfrak{r}$ is not equal to any of the numbers in the original list, since it differs from each number in at least one digit. 
However, this contradicts our assumption that all the real numbers in $(0,1)$ are listed in the matrix. Therefore, the assumption that there is a one-to-one correspondence between the set of real numbers in $(0,1)$ and the set of integers is false. Hence, the set of real numbers in $(0,1)$ is uncountable, and there are more real numbers than there are integers.

Using the above proof, we can immediately conclude  
that it is impossible to use an alphabet with a finite number of characters to express every real number as a finite-length string, which sets them apart from integers and rationals. For example, in binary notation, any integer can be represented with only three characters, `0', `1', and `--' in finite length, as demonstrated by 11 and --1011. Similarly, any rational number can be expressed as a finite-length string of four characters, `0', `1', `--', and `/', such as 11/10 and --101/11. However, this is not the case for real numbers.  It is true that some irrational numbers 
can be represented as finite-length strings with an expanded alphabet. For instance, using three additional characters `(,),\^{}' (i.e., left and right parentheses and an upper cusp), we can represent the irrational number $\sqrt{3}$ as 3\^{}(1/2), a string of length 7. Similarly, $5^{1/3}$ can be expressed as 5\^{}(1/3) using the same alphabet. By adding more characters into the alphabet, we can express more irrational numbers as strings of finite length. For example, we can express $\sqrt{\pi}$ as $\pi$\^{}(1/2) by adding $\pi$ to the alphabet. However, one can prove that no alphabet of finite size can express every irrational number as a finite-length string.

To demonstrate this, we assume the opposite that one can use a finitely large alphabet to express every irrational number as a finite-length string. We can then represent each character in the alphabet as a binary integer, just as in a computer. Consequently, each irrational number is expressed as a finitely large integer, which can be arranged in a one-to-one correspondence with the natural numbers. However, this contradicts Cantor's proof that there is no one-to-one correspondence between the natural numbers and the real numbers, especially the irrational ones. Therefore, we cannot express every irrational number as a finite-length string with an alphabet of finite size. This conclusion may seem obvious, but it is profound because it applies to any set with an uncountably infinite number of elements. In other words, it is impossible to build an alphabet of finite size that expresses any element of this set as a finite-length string. This conclusion is crucial for the following discussion.

Let us consider the set $\Sigma$ of all mathematical proofs, which includes existing mathematical proofs and those that will be discovered in the infinitely long future. To examine the characteristics of mathematical proofs, 
we can use Cantor's proof for real numbers as an example. Cantor's proof begins by assuming that 
real numbers are countable. According to existing mathematical knowledge, countable sets can always be labeled with natural numbers. Therefore, all real numbers can be written down as an infinite series $r_1,r_2,r_3\cdots,r_n,\cdots$. Cantor then constructs a matrix using this series, as shown in equation (\ref{eq:cantorm}).
Finally, a new real number $\mathfrak{r}$ is created by flipping the diagonal elements of this matrix, which contradicts the initial assumption that all the real numbers are already listed in the matrix. 
The real numbers are thus proved to be uncountable. This proof has an interesting feature: although it involves an infinite-dimensional matrix (\ref{eq:cantorm}), the total length of the proof is still finite.
This is because Cantor does not need to know the details of each real number in the proof, such as whether the millionth digit of the second real number $r_2$ is 0 or 1. 
These details are irrelevant to the proof, as Cantor is only concerned with one basic feature of real numbers in (0,1): any array of 0s and 1s corresponds uniquely to one such real number. 
This feature ensures that the new number $\mathfrak{r}$ constructed is a real number in (0,1).

From the discussion above, it follows that every mathematical proof can be represented by a finite-length string. The alphabet used in a mathematical proof includes all the characters in the language used (such as the 26 letters of the English alphabet and their punctuation) and all the symbols in mathematics, and this alphabet is certainly finite. As a result, the set $\Sigma$ of all mathematical proofs is a countably infinite set. This conclusion reflects a fundamental physical fact: any mathematical proof is carried out by mathematicians or computers, or both, and each of these entities is a finite physical system that can only use a finite number of letters and symbols, perform a finite number of transformations and operations, and take finite steps of derivation.

Consider the set of all mathematical propositions, denoted by $\Omega$. Although each proposition appears to be a finite-length string with a finite alphabet, this set is surprisingly uncountable. For instance, one can construct a proposition for each real number. To illustrate, we define the notion of embedding a number $x$ in $\pi$, which means that all the digits of $x$ appear in order in the decimal expansion of $\pi$. For example, the number $x=1.234567$ can be embedded in $\pi$ as
\begin{equation}
3. \underline{\bm 1}4159\underline{\bm 2}65\underline{\bm 3}5 89793238\underline{\bm 4}6 2643383279 
\underline{\bm 5}028841971 \underline{\bm 6}93993\underline{\bm 7}510...
\end{equation}
For any real number $x$ and every natural number $n$, we define $R^{(n)}(x)$ as the operation that repeats each digit of $x$ $n$ times. For example, 
$R^{(1)}(23456)=23456$, $R^{(3)}(1.36)=111.333666$. For $n>1$, $R^{(n)}(x)$ is embedded in $\pi$ if each  digit of $x$ in its repeated form appears in order
in $\pi$.  For example, $R^{(2)}(1414)=11441144$ is embedded in $\pi$, 
\ba
&&3.1415926535 8979323846 2643383279 5028841971 6939937510 \nonumber\\
&&5820974944 5923078164 0628620899 8628034825 342\underline{\bm {11}}70679 \nonumber\\
&&8214808651 3282306647 
0938\underline{\bm{44}}6095 5058223172 5359408128 \nonumber\\
&&48\underline{\bm{11}}174502 8410270193 8521105559 6\underline{\bm{44}}6229489 ...
\ea
For every real number $x$ and every natural number $n$.
we can construct a mathematical proposition $E^{(n)}(x)$, which
states that $R^{(n)}(x)$ is embedded in $\pi$. As all mathematical propositions $E^{(n)}(x)$ belong to  $\Omega$, 
the set $\Omega$ is uncountably infinite.

As the set of all mathematical propositions $\Omega$ has uncountably infinitely many elements, while the set of all mathematical proofs $\Sigma$ is countably infinitely many, it is impossible to establish a one-to-one correspondence between these two sets. Therefore, there are always propositions in $\Omega$ that cannot be proved or disproved, and in fact, an infinite number of such propositions exist. For example, the proposition $E^{(1)}(\sqrt{2})$ is very likely to be true, but it seems impossible to prove it because we would need to compare two infinitely long and irregular sequences of numbers. Neither the human brain nor any computer can record an infinite sequence of numbers. On the other hand, we cannot completely rule out the possibility of falsifying this proposition. If, by calculation or other means, we find that the number 9 no longer appears after the $2^{200}$th decimal place in $\pi$, and $\sqrt{2}$ has the number 9 appearing after the $2^{200}$th decimal place or higher, then the proposition $E^{(1)}(\sqrt{2})$ is wrong. I expect that most propositions $E^{(n)}(x)$ are impossible to prove or falsify.

In conclusion, while the preceding discussion lacks mathematical rigor, it reveals a fundamental truth: mathematicians and computers are finite physical entities that are subject to limitations. They can only use a finite number of symbols, perform a finite number of operations, and take a finite number of steps in a proof. As a result, there will always be mathematical propositions that are beyond the reach of human or computational proof. This realization demonstrates that mathematics is inherently physical, and that the limitations of our physical existence necessarily constrain our ability to fully comprehend it. However, they also ensure that there will always be new and challenging problems for mathematicians to tackle. So, far from rendering mathematicians unemployed, the incompleteness of mathematics guarantees that their work will continue to be essential and endlessly fascinating.

The above discussion and conclusion reflect the essence of G\"odel's theorem, even though they are not particularly rigorous. G\"odel very creatively encoded all mathematical expressions in natural numbers, including both mathematical formulae like $1=1$ and meta-mathematical expressions like "x is a real number." Finally, using Cantor's diagonal argument, he proved what is now known as G\"odel's incompleteness theorem. Due to space constraints, we cannot delve into the details of the proof here. Interested readers can refer to G\"odel's original paper \cite{godel} or books on the theorem \cite{nagel}.

Interestingly, Turing approached the undecidability of mathematical propositions from the perspective of computing and proposed an equivalent problem, known as the halting problem \cite{chuang,turing}. Given an input, a program $\mathfrak{p}$ can either halt after completing a computation or run into an infinite loop. The halting problem asks whether there exists a program $\mathfrak{P}$ that can determine, in a finite amount of time, whether any given program $\mathfrak{p}$ will halt or loop infinitely.

To demonstrate the impossibility of such a program, we consider a special class of programs $\Xi$ that take only positive integers as inputs. There are infinitely many programs in $\Xi$, which can either be countable or uncountable. 
Since whether a program $\mathfrak{p}$ in $\Xi$ halts on an input is essentially a mathematical proposition, 
if $\Xi$ were uncountable,  then the existence of a program $\mathfrak{P}$ that could determine whether any given program $\mathfrak{p}$ halts or loops forever would result in an uncountable number of mathematical proofs. This is contradictory to the previous conclusion that the set of all mathematical proofs, $\Sigma$, is countable. Therefore, we have to assume the second possibility, that $\Xi$ has a countably infinite number of elements. We can denote the elements of $\Xi$ as $\mathfrak{p}_1,\mathfrak{p}_2,\mathfrak{p}_3,\cdots,\mathfrak{p}_n,\cdots$, and for a given input $m$, the program $\mathfrak{p}_n$ either halts or loops forever.

We further assume that there exists a program $\mathfrak{P}$ that can determine whether each program $\mathfrak{p}_n$ in $\Xi$ halts or loops forever on input $m$. Specifically, if $\mathfrak{p}_n$ halts on input $m$, $\mathfrak{P}$ outputs 1, and if $\mathfrak{p}_n$ loops forever on input $m$, $\mathfrak{P}$ outputs 0. 
These outputs form a matrix (see, for example, Eq.(\ref{eq:halt})), which we can use to construct a new program $\tilde{\mathfrak{p}}$ using diagonal argument. The program $\tilde{\mathfrak{p}}$ halts on input $n$ if $\mathfrak{P}(\mathfrak{p}_n,n)=0$, and loops forever on input $n$ if $\mathfrak{P}(\mathfrak{p}_n,n)=1$. This program is different from all programs in $\Xi$ because $\mathfrak{P}(\tilde{\mathfrak{p}},n)\neq \mathfrak{P}(\mathfrak{p}_n,n)$. This contradicts the assumption that we have already listed all the elements in $\Xi$ as $\mathfrak{p}_1,\mathfrak{p}_2,\mathfrak{p}_3,\cdots,\mathfrak{p}_n,\cdots$. Therefore, the assumption is wrong, and the program $\mathfrak{P}$ does not exist.

\begin{equation}
\label{eq:halt}
\begin{array}{c|cccccccccc}
m   &1 & 2 & 3 & 4 & 5 & 6 & 7 & 8 &9 &\cdots\\
\hline
\mathfrak{p}_1=&\underline{0}&1&1&0&1&0&1&0&0& \cdots\\ 
\mathfrak{p}_2=&1& \underline{0}&1&1&1&0&1&1&0& \cdots\\
\mathfrak{p}_3=&0&0&\underline{1}&1&1&1&0&0&0& \cdots\\ 
\mathfrak{p}_4=&0&1&1&\underline{1}&0&0&1&1&0& \cdots\\ 
\mathfrak{p}_5=&1&0&1&0&\underline{0}&1&0&1&0& \cdots\\ 
\mathfrak{p}_6=&0&0&1&0&0&\underline{1}&0&0&0& \cdots\\ 
\mathfrak{p}_7=&0&1&1&1&0&1&\underline{0}&0&0& \cdots\\ 
\mathfrak{p}_8=&1&1&1&1&0&1&0&\underline{1}&0& \cdots\\ 
\mathfrak{p}_9=&1&1&1&0&0&1&1&1&\underline{1}& \cdots\\ 
\vdots &\vdots &\vdots &\vdots &\vdots &\vdots &\vdots &\vdots &\vdots &\vdots &\ddots
\end{array}
\end{equation}

Although the above argument on the halting problem differs from the commonly seen proof \cite{chuang}, its essence is the same. The proof presented here demonstrates that the number of programs is uncountably infinite, and that this is the fundamental reason why it is impossible to determine whether any program halts or not. This limitation is analogous to our inability to prove or falsify all mathematical propositions. By highlighting this connection, the proof provides a deeper understanding of the limits of computation and mathematical knowledge.

\section{Discussion and Perspectives}
\label{sec:diss}
Galileo once famously declared in his work \textit{The Assayer} \cite{assayer}, ``The universe is a great book open for the perusal of men. To understand the philosophy of it, you must first learn its language. The language of the book is mathematics, and the characters it uses are triangles, circles and other geometric figures. Without knowing these, you cannot understand a single word in the book, you can only wander in a dark maze." In this quote, Galileo was referring to what we now call science, or more specifically, physics. His statement was bold and visionary because at that time, there were only a few physical phenomena, such as free fall and the single pendulum, that could be described with mathematical precision. However, the development of physics in the centuries since Galileo's time has confirmed his vision, and mathematics has become an indispensable tool for physicists. Not only is mathematics the language in which physicists describe the universe, but it has also profoundly influenced the way physicists approach and solve problems. For example, theoretical physicists never hesitate to discuss idealized situations such as massive particles without size 
and experiments without noise. 

Physics and mathematics have a symbiotic relationship where both fields mutually influence each other. Physics has inspired and provided material for mathematical thinking, such as the equations of hydrodynamics and models of statistical mechanics. On the other hand, mathematics has played a crucial role in the development of modern physics, serving as a language to describe and formulate physical theories.
However, the influence of physics on mathematics goes beyond just inspiration and material. What is discussed in this paper highlights a fundamental principle that mathematics is ultimately influenced by the underlying physical entities. For example, as mathematicians and/or computers are of finite physical resources,   there are mathematical statements that cannot be proven within a formal system, and by  exploiting the principles of quantum mechanics, quantum computers may dramatically improve the efficiency of   computation. 

It can be confidently predicted that principled influences will continue to shape the future in increasingly diverse ways. One possibility is the development of computers with greater processing power than even the brightest human brains. These computers, as new physical entities, will be capable of proving mathematical propositions that humans are unable to prove. This will lead to the discovery of many new mathematical relationships that the human brain may struggle to comprehend, much like how most people today cannot fully grasp the complex mathematics developed by leading mathematicians.

Another potential development is the creation of a new system of dynamics by physicists, such as in their efforts to unify gravity and quantum mechanics. Similar to how quantum dynamics has given a new meaning to Hilbert space, this new dynamical system may provide new connotations to well-established mathematics and offer a new kind of information. Computers that leverage this information will have greater power and be able to solve problems that were previously unsolvable. In fact, a recent paper  \cite{HWW} suggests a possible new kind of information and proposes a new computer model that is more powerful than even a quantum computer.

The soaring eagle appears to effortlessly overcome the invisible force of gravity, flying without restraint. However, the truth is that it can never be entirely free from its influence. Similarly, while mathematics serves as the language of physics and can often develop independently, it can never transcend physics completely. The influence of underlying physics on mathematics may be subtle, but it can be very powerful.

\acknowledgments
 The work was supported by the 
 National Key R\&D Program of China (No.~2018YFA0305602) and
 the National Natural Science Foundation of China (Grant No. 11921005).

%
%

\end{document}